\documentclass[a4paper,10pt]{article}
\usepackage[sort]{cite}
\usepackage{bookmark}
\usepackage[utf8]{inputenc} 
\usepackage[T1]{fontenc}    
\usepackage{url}            
\usepackage{booktabs}       
\usepackage{amsfonts}       
\usepackage{amsmath}
\usepackage{amsthm}
\usepackage{amssymb}
\usepackage{mathrsfs}
\usepackage{graphicx} 
\usepackage{tabularx}
\usepackage{tikz-cd}
\usepackage{tikz}
\usepackage{proof}
\usepackage{caption}
\usetikzlibrary{matrix}
\usepackage{nicefrac}     
\usepackage{multirow} 
\usepackage{float}
\usepackage{microtype}      
\usepackage{doi}
\usepackage[nottoc]{tocbibind}
\usepackage{ragged2e}
\usepackage{float}
\usepackage{chngcntr}
\usepackage{enumerate}
\usepackage{fullpage}
\usepackage{blindtext}
\usepackage{stmaryrd }
\usepackage{nomencl}
\usepackage{etoolbox}
\usepackage[draft]{fixme}

\newtheorem{theorem}{Theorem}[section]
\newtheorem{corollary}{Corollary}[theorem]
\newtheorem{lemma}{Lemma}[section]

\newtheorem{proposition}{Proposition}[section]

\newtheorem{remark}{Remark} [section]

\newtheorem{definition}{Definition}[section]

\newtheorem {conjecture}{Conjecture}

\newcommand\gdl[1]{\mathbf {G}_{#1}}

\newcommand\IL{\ensuremath{\mathbf{IL}}}

\newcommand\bbR{\mathbb{R}}

\newcommand\del{{\mathop{\triangle}}}
\newcommand\imm{\ensuremath}
\newcommand\limp{\to}
\newcommand\impl{\limp}

\newcommand\qe[1]{\exists #1}
\newcommand\qa[1]{\forall #1}

\newcommand\axlin{{\ensuremath{\hbox{\scshape lin}}}}
\newcommand\axqs{{\ensuremath{\hbox{\scshape qs}}}}

\newcommand\axiso{{\ensuremath{\hbox{\scshape iso}_0}}}
\newcommand\axisoi{{\ensuremath{\hbox{\scshape iso}_1}}}
\newcommand\axfinn{{\ensuremath{\hbox{\scshape fin}(n)}}}

\newcommand\axdi{{\ensuremath{\axdelta1}}}
\newcommand\axdii{{\ensuremath{\axdelta2}}}
\newcommand\axdiii{{\ensuremath{\axdelta3}}}
\newcommand\axdiv{{\ensuremath{\axdelta4}}}
\newcommand\axdv{{\ensuremath{\axdelta5}}}

\newcommand\Proves\Rightarrow
\newcommand\nProves\nRightarrow

\newcommand\I{\imm{\mathcal I}}

\newcommand\lequiv{\leftrightarrow}

\newcommand\nmodels{\mathrel{\nvDash}}

\newcommand\axdelta{\mathnormal{\mathrm{\Delta}}}

\newcommand\entails{\mathrel{\models}}
\newcommand\nentails{\mathrel{\nmodels}}

\relax

\newcommand{\shorttitle}{\@title}

\renewenvironment{abstract}
{
  \centerline
  {\large \bfseries \scshape Abstract}
  \begin{quote}
}
{
  \end{quote}
}

\def\keywordname{{\bfseries \emph{Keywords}}}%
\def\keywords#1{\par\addvspace\medskipamount{\rightskip=0pt plus1cm
\def\and{\ifhmode\unskip\nobreak\fi\ $\cdot$
}\noindent\keywordname\enspace\ignorespaces#1\par}}

\title{Gödel logics: Prenex fragments \thanks{Research supported by FWF grant P 36571}  }

\author{ 
{
\hspace{1mm}Matthias Baaz} \\
Technische Universität Wien\\
Institut für Algebra und 
Diskrete Mathematik \\
A–1040 Vienna, Austria\\
	\texttt{Baaz@logic.at} \\
 {
 \hspace{1mm} Mariami Gamsakhurdia} \\
	Technische Universität Wien\\
Institut für Algebra und 
Diskrete Mathematik\\ 
A–1040 Vienna, Austria\\
	\texttt{mariami.gamsakhurdia@tuwien.ac.at} \\
}

\date{}

\begin{document}
\maketitle
\begin{abstract}
In this paper, we provide a complete classification for the first-order Gödel logics concerning the property that the formulas admit logically equivalent prenex normal forms. We show that the only first-order Gödel logics that admit such prenex forms are those with finite truth value sets since they allow all quantifier-shift rules and the logic \(G_\uparrow\) with only one accumulation point at $1$ in the infinite truth values set. In all the other cases, there are generally no logically equivalent prenex normal forms. We will also see that \(G_\uparrow\) is the intersection of all finite first-order Gödel logics.\\
The second part of this paper investigates the existence of effective equivalence between the validity of a formula and the validity of some prenex normal form. The existence of such a normal form is obvious for finite valued Gödel logic and \(G_\uparrow\). Gödel logics with an uncountable truth value set admit the prenex normal forms if and only if every surrounding of \(0\) is uncountable or \(0\) is an isolated point. Otherwise, uncountable Gödel logics are not recursively enumerable, however, the prenex fragment is always recursively enumerable. Therefore, there is no effective translation between the valid formula and the valid prenex normal form. However, the existence of effectively constructible validity equivalent prenex forms for the countable case is still up for debate.
\end{abstract}


\begin{section}{Introduction}\label{intro}
 \vspace{-0.5ex} 

 \subsection{Prenex Fragments}\label{1.1}

The existence of an equivalent prenex normal form is one of the oldest recognized characteristics of classical predicate logic: to every formula corresponds an equivalent prenex formula. The prenex formulas have many features that usual formulas do not have; One can skolemize, in certain cases, it admits Herbrand's theorem or some infinite form of it. This concept of prenexation is not as simple as one might think, because the prenex forms can be linguistically different. There might exist prenex formulas, although not all quantifier shift rules are valid. A natural example is to have some fragment of non-classic logic, where every sentence is either true or false; Then we have prenex forms and prenexation has nothing to do with quantifier shifts.\\ 
The expressive power of prenex fragments is easy to see in classical logic because it coincides with the whole logic; It is also easy to see for intuitionistic logic since the validity of prenex formulas is decidable. Prenex fragments in non-classical logics normally do not coincide with the logic, but they are easier to handle syntactically and semantically. 
Typically, the prenex fragment has a structure that can be analyzed by algebraic tools. As a result, even though the logic might not interpolate (such as first-order S5), the prenex fragment does \cite{Baaz-Zach:HypersequentIntuitionisticFL},\cite{Davey:Subvarieties}.
Since Gödel logics are intermediary logics, the expressibility of prenex fragments is in doubt.\\ 
Of course, it is easy to see that one can not hope to obtain prenex normal forms in the usual sense in Gödel logics. Note that the absence of the quantifier-shift rules does not guarantee that no prenex normal form exists. However, proving that such prenex forms do not exist is more difficult because the formula may have a completely different nature.\\
Usually, quantifier shifting rules are applied to construct these prenex normal forms.
Since the skolem functions relate to quantifier shifting, the prenex fragments of Gödel logics admit skolemization while the logics themselves usually do not. 
Furthermore, the prenex fragments admit Herbrand’s theorems or approximative versions (for example, $G_\uparrow$). In addition, they provide general normal form theorems that demonstrate the lower bound of the complexity of particular Gödel logics.
The length of direct, i.e. cut-free proofs and Herbrand disjunctions, however, is greatly influenced by the shifting of quantifiers \footnote{This is where Herbrand's mistake in the original proof of Herbrand's theorem came from: he believed that the quantifier shifting did not alter the Herbrand disjunction, and his reasoning was limited to prenex forms \cite{Baaz-Veith:QuantifiedPropositionalGodel}.}.\\
 Interestingly, the 1-satisfiability of prenex formulas in Gödel logics is a classical satisfiability (unlike validity). This is important for database-like structures.
 For ontologies, it is important to prove that they are consistent (1-satisfiable). The fact that rules subsume another one is of minor importance. The many-valued structure being non-contradictory can be, therefore, reduced to a classical satisfiability problem. \\
As a first example note that prenexation does not work for \(G_{[0,1]}\) when 0 is not isolated, since the formula 
    \((\neg \forall x A(x)\wedge  \forall x \neg \neg A(x))\)
does not admit a prenex normal form.
This result can be extended to all Gödel logics where there is at least one accumulation point from above, even if it is not 0. \\
The logic might not admit prenex forms in the original language, but it might admit prenex normal forms in an extended language:

\textbf{Example}\label{example} (First-order Gödel on $[0,1]$ logic with propositional quantifiers):\\
 There are only three quantifier-shift rules that are not intuitionistically valid in first-order Gödel logics
 \begin{align*}
  (S1)  \quad  \forall x(A(x)\vee B)\supset(\forall xA(x) \vee B) \\
(S2)   \quad ( B\supset \exists xA(x))\supset(\exists x B\supset \vee A(x))\\
(S3) \quad  (\forall x A(x)\supset B)\supset(\exists x(A(x) \supset B)  
\end{align*}
(\(x\) is not free in \(B\)). \\
Using propositional quantifiers prenexation is possible in $G_{[0,1]}$ as the following quantifier shifts are valid 
 \begin{align*}
  (Q1)  \quad  \forall xA(x)\supset B\lequiv \forall Z\exists x(A(x) \supset Z\vee Z\supset B) \\
(Q2)   \quad ( A\supset \exists xB(x))\lequiv\forall Z\exists x(A \supset Z\vee Z\supset B(x)).
\end{align*}
For the right direction use the validity of $A\supset B\supset((A\supset C)\vee (C \supset B))$ for all formulas $C$, for the left direction use the validity of the quantified propositional expression of the Takeuti-Titani Rule $\forall Z(A\supset Z\vee Z\supset A)\supset (A\supset B)$. All other classical quantifier shift rules are valid in the extended language.


In this paper, we study two different problems:
\begin{enumerate}
    \item Logical equivalence
    \item Effective validity equivalence
\end{enumerate}
We give a comprehensive classification of first-order G\"odel logics after the existence of the equivalent prenex normal forms.

\subsection{First-order Gödel Logics}\label{1.2}
\vspace{-0.5ex} 

We begin with a preliminary discussion of the syntax and semantics of first-order Gödel logics, followed by some of the more interesting special cases and their relationships.\\
First-order Gödel logics are a family of many-valued logics where the truth values set (known also as \emph{G\"odel set}) \(V\) is closed\footnote{Gödel logics are usually defined using the single truth value set $[0,1]$. For propositional
logic the choice of any infinite subset of $V\subseteq [0,1]$ leads to the same propositional
logic (set of tautologies). In the first order case, where quantifiers will be interpreted as infima and suprema, a closed subset of $[0,1]$ is necessary.} subset of the full \([0,1]\) interval that includes both $0$ and $1$. 
We fix a standard first-order language $\mathcal{L}$ with finitely or countably many predicate symbols $P$ and finitely or countably many function symbols $f$ for every finite arity $k$.

\begin{definition}\label{d - 1.1} First-order Gödel logic \(G_V\) is a logic defined over the \emph{Gödel set} \(V\) (any closed set of real numbers, $V \subseteq [0,1]$ containing $0$ and $1$) given by the following evaluation function \(\mathcal{I}\) on $V$
\begin{align*}
   (1) & \quad \mathcal{I}(\bot)= 0\\
   (2) & \quad \mathcal{I}(A \wedge B)= min \{\mathcal{I}(A), \mathcal{I}(B)\}\\
   (3) & \quad \mathcal{I}(A \vee B)= max \{\mathcal{I}(A), \mathcal{I}(B)\} \\
   (4) & \quad  \mathcal{I}(A \supset B) = \begin{cases}\I(B) & \text{if\/ $\I(A) > \I(B)$,} \\
               1     & \text{if\/ $\I(A) \le \I(B)$.}\end{cases}   \\       
 (5) & \quad \mathcal{I}(\forall x A(x))= \text{inf}\{\mathcal{I}( A(u))\: u\in U\} \\
 (6) & \quad\mathcal{I}(\exists x A(x))= \text{sup} \{\mathcal{I}( A(u))\: u\in U\} 
\end{align*}
    
\end{definition}

\begin{remark}\label{r - 1.1}
Note that the fourth operator is characteristic operator of Gödel logics, so-called Gödel conditional. 
\end{remark}

\begin{remark}
We use the abbreviation $A<B$ for \((A\supset B)\supset A\), which means that the value of \(A\) is bigger than the value of \(B\) with the exception that both take the value $1$  
and $A\lequiv B$ for $ (A\supset B) \land (B\supset A)$.
\end{remark}

Properties of first-order Gödel logics are based on the order-theoretic structure of their set of truth values rather than the truth values themselves. 
In particular, different logics are distinguished by the number and type of accumulation points and their respective Cantor–Bendixon ranks \cite{Baaz:Delta},\cite{Beckmann-Goldstern-Preining:Fraisse},\cite{Kechris:DescriptiveSetTheory}. This raises the question of an intentional\footnote{ We use intentional in the sense that logical properties (e.g., set of validities, entailment relation, set of satisfiable formulas) change when the characterising object changes.} characterization of Gödel logics. While it is technically convenient to represent Gödel logics via their set of truth values, it has the disadvantage that uncountably many truth value sets have indistinguishable logical properties. 
For example, consider 
\begin{center}
 \(V_1 = \{1/n : n\in N^+\} \cup \{0\}\) \\
\(V_2 = \{1/2n : n\in N^+\}\cup\{0\}\).   
\end{center} The logics \(G_{V_1}\) and \(G_{V_2}\) have exactly the same properties.
This leads to the representation of Gödel logics as a set of valid sentences.
\begin{definition} \label{d - 1.2} We define the set of valid formulas \(G_V\) (referred to as Gödel logic) as the set of formulas \(A\) of \(\mathcal{L}\) such that the evaluation of \(A\) is 1 for all 
\(V\)-interpretations \(\mathcal{I}\).
 \end{definition}
While focusing on validities, we must address insufficiencies with respect to satisfiability and entailment. It is unclear how to describe the sets of 1-satisfiable sentences in these logics. Furthermore, whereas there are uncountable many entailment relations, there are only countable many sets of validities [\cite{Baaz-Veith:Interpolation},\cite{Baaz-Zach:Compact}]. Even the 3-valued Gödel logic \(G_3\) corresponds to uncountably many Gödel sets
\(\{0, c, 1\}\) for any \(c \in(0, 1)\). However, this definition would be problematic if different sets
of satisfiable formulas correspond to identical Gödel logics depending on the topological structure. This is prevented by a functional dependency between the set of sets of valid formulas and the set of sets of $1$-satisfiable formulas \cite{Baaz-Preining:classificationoffirstorder}.
As a generalisation of classical satisfiability, we introduce the
following concepts:

\begin{definition}[satisfiability]\label{d - 1.3} The formula in Gödel logic is \emph{1-satisfiable ($>0$ satisfiable)} if there exists at least one interpretation
that assigns 1 ($>0$) to the formula.  
\end{definition}
\begin{definition}[Validity]\label{d - 1.4}
The formula in Gödel logic is \emph{valid} (\emph{$>0$-valid}) if the formula evaluates to 1 ($>0$) under every interpretation. 
\end{definition}
An interesting aspect to note is that in most investigations of non-classical logics, validity is prioritised over $1$-satisfiability. These two are, of course, dual in classical logic, in the sense that $A$ is valid iff $\neg A$ is not satisfiable. This duality does not hold in many-valued logics: Consider, for example, the first-order Gödel logic based on $\{1 - 1/n :n \in N\} \cup\{1\}$. There, validity is not recursively enumerable, but both the complement of $1$-satisfiability and the complement of  $>0$ -satisfiability are recursively enumerable (see \cite{Baaz-Preining:classificationoffirstorder} Lemma 37 and 38).\\ 
Equivalence between formulas in G\"odel logic are defined in the usual way,i.e., two formulas are equivalent if they are equal under all interpretations.

\begin{definition}\label{d - 1.5}
For G\"{o}del logics there is an
asymmetry between~$0$ and~$1$, this is because~$0$ can be distinguished from other values (by using the negation), while~$1$ cannot be
distinguished. The reason is that all connectives and quantifiers are
continuous at~$1$. To overcome this asymmetry the absoluteness operator~$\del$ has
been introduced in~\cite{Baaz:Delta}.
\begin{align*}    
 \I(\del A) = \begin{cases} 1 & \text{if $\I(A)=1$,}\\ 0 &\text{otherwise}\end{cases}
\end{align*}
\end{definition}
We consider Gödel logics with or without an absoluteness  operator~$\del$ and denote by $G_V^\del$ or $G_V$, respectively.





We recall several definitions from topology and descriptive set theory.

\begin{definition}\label{d - 1.6}(Limit point, perfect space, perfect set).
A \emph{non-isolated point} of a topological space is a point \(x\) such that for every open neighbourhood \(U\) of \(x\) there exists a point \(y\in U\) with \(y\neq x\). A \emph{limit point} of a topological space is a point \(x\) that is not isolated.  A space is \emph{perfect} if all its points are limit points.
A set \(P\subseteq R\) is \emph{perfect} if it is closed and together with the topology induced from \(R\) is a perfect space. 
\end{definition}

It is obvious that all (non-trivial) closed intervals are perfect sets, as
well as all countable unions of (non-trivial) intervals. But all these sets
generated from closed intervals have the property that they are `everywhere
dense,' i.e., contained in the closure of their inner component. 

Let us list all the occurring axioms and rules:
\begin{align*}
  \hbox{(I1)} &\quad (A\limp B) \limp ((B \limp C) \limp (A\limp C))\\
  \hbox{(I2)} &\quad A\lor A\limp A, A\limp A\land A  \\
  \hbox{(I3)} &\quad A\limp A\lor B, A\land B\limp A\\
  \hbox{(I4)} &\quad A\lor B\limp B\lor A, A\land B\limp B\land A \\
  \hbox{(I5)} &\quad (A\limp B)\limp (C\lor A\limp C\lor B)\\
  \hbox{(I6)} &\quad (A\land B\limp C) \limp (A\limp(B\limp C))\\
  \hbox{(I7)} &\quad (A\limp (B\limp C)) \limp (A\land B\limp C)\\
  \hbox{(I8)} &\quad \bot\limp A \\
  \hbox{(I9)} &\quad \qa xA(x) \limp A(t)\\
  \hbox{(I10)} &\quad A(t)\limp \qe xA(x) \\
  \intertext{Additional axioms for~$\del$:}
  \axdi &\quad \del A\lor \neg\del A\\
  \axdii &\quad \del (A\lor B) \limp (\del A \lor \del B)\\
  \axdiii&\quad \del A \limp A\\
  \axdiv &\quad \del A \limp \del\del A\\
  \axdv   &\quad \del (A\limp B)\limp(\del A\limp\del B)
  \intertext{Additional axioms for Gödel logics:}
  \axqs  &\quad \qa x(B^{(x)}\lor A(x)) \limp (B^{(x)}\lor\qa xA(x)) \\
  \axlin &\quad (A\limp B)\lor (B\limp A)\\
  \axiso &\quad \qa x\neg\neg A(x) \limp  \neg\neg\qa xA(x)\\
  \axisoi &\quad \del \qe x A(x)\limp \qe x \del A(x)\\
  \axfinn &\quad (\top\limp p_1) \lor (p_1\limp p_2)\lor \ldots \lor (p_{n-2}\limp p_{n-1})\lor (p_{n-1}\limp\bot)
\end{align*}
and the set of rules is
\begin{align*}
  \hbox{(MP)} &\quad \vcenter{\infer{B}{A & A\limp B}} &
  \hbox{(I$\del$)} & \quad \vcenter{\infer{\del A}{A}} \\               
  \hbox{(Q$\forall$)} &\quad \vcenter{\infer{B^{(x)}\limp\qa xA(x)}{B^{(x)}\limp A(x)}} &
  \hbox{(Q$\exists$)} &\quad \vcenter{\infer{\qe xA(x)\limp B^{(x)}}{A(x)\limp B^{(x)}}}
\intertext{(where $B^{(x)}$ means that $x$ is not free in $B$).}
\end{align*}
The names of the axioms can be explained as follows: $\axqs$ stands
for `quantifier shift', $\axlin$ for `linearity', $\axiso$ for
`isolation axiom of $0$', $\axisoi$ for `isolation axiom of $1$',
and $\axfinn$ for `finite with $n$ elements'.

The axioms for isolation have been chosen to be purely universally
(case~$0$) and purely existentially (case~$1$). $\axiso$ expresses
that if all instances are larger than~$0$, then also the
infimum. $\axisoi$ expresses that if the supremum is~$1$, then there
is also an instance that is~$1$. 
There are alternative axioms that are equivalent:
\begin{align*}
  \axiso'  &\quad \neg\qa xA(x)\limp\qe x\neg A(x)\\
  \axisoi' &\quad \forall x\neg \del A(x)\limp\neg\del\exists xA(x)
\end{align*}


\subsection{Relationships between first-order Gödel logics }\label{1.3}

\vspace{-0.5ex}
The relationships between finite and infinite valued propositional Gödel logics are well understood. Any choice of an infinite set of truth values results in the same propositional Gödel logic. In the first-order case, the situation is more interesting.
We establish several important results regarding the relationships between various first-order
Gödel logics. We first consider several ‘prototypical’\footnote{This is because logic is defined extensionally as the set of formulas valid in this truth value set, so the Gödel logics on different truth value sets may coincide.} Gödel sets:\\
\begin{center}
\(V_{[0,1]}=[0,1]\)\\
\(V_\downarrow=\{0\}\cup \{1/k : k\geq 1\}\)\\
\(V_\uparrow=\{1\}\cup \{1 - 1/k : k\geq 1\}\)\\
 \(V_m=\{1\}\cup \{1 - 1/k : 1\geq k\geq m-1\}\)  
\end{center}
The corresponding Gödel logics are \(G_{[0,1]}\), \(G_\downarrow\), \(G_\uparrow \), \(G_m\). \(G_{[0,1]}\) is \emph{standard} Gödel logic.
The logic \(G_\downarrow\) also turns out to be closely related to some temporal logics.\\
Below we present relationships between some first-order Gödel logics \cite{Baaz-Leitsch-Zach:TimeGaps},\cite{Baaz-Leitsch-Zach:FirstOrderGoedel}.

\begin{proposition}\label{p - 1.1}
Whenever $V\subseteq V'$ then $G_{V'} \subseteq G_V$.
\end{proposition}
\begin{proof}
If $V \subseteq V'$ are both Gödel sets, and $\mathcal{I}$ is an interpretation into $V'$, then $\mathcal{I}$ can also be seen as an interpretation into
$V$, and the values $\mathcal{I}(A)$, computed recursively using (1) – (6) in Definition 1.2, do not depend on whether we consider $\mathcal{I}$ as a $V'$-interpretation or a $V$-interpretation. Consequently, if $V' \subseteq V$, there are more interpretations into $V$ than into $V'$, hence if $\Gamma\models_{V'} A$ then also $\Gamma\models_V A$ and $G_{V'}\subseteq G_V $.
\end{proof}
The result can be generalized to the embeddings between Gödel sets.
\begin{definition}[g-embedding]\label{d - 1.7}
  A \emph{g-embedding} $h$ from a Gödel set~$V_1$ to a Gödel set~$V_2$
  is an embedding $h: V_1\to V_2$ which is continuous (in the
  usual topology of closed sets on~$\bbR$), strictly monotone on
  $V_1$, and preserves~$0$ and~$1$.
\end{definition}

\begin{proposition} \label{p - 1.2}\cite{Baaz-Leitsch-Zach:FirstOrderGoedel} The following containment relationships hold:\\
1) \(G_m\supsetneq G_{m+1}\)\\  
2) \(G_m\supsetneq G_\uparrow \supsetneq G_{[0,1]}\)   \\
3) \(G_m\supsetneq G_\downarrow \supsetneq G_{[0,1]}\)

\end{proposition}

\begin{proof}
The only challenging part is demonstrating strict containment.  Note that the $(Fin)$ $(\top\supset A_1)\vee(A_1\supset A_2)\vee (A_2\supset A_3)\vee \dots \vee (A_{m-2}\supset A_{m-1})\vee (A_{m-1}\supset \bot)$ is valid in 
\(G_m\) but not in \(G_{m+1}\). Consider the formulas 
\begin{center}
    (1) $\quad $ \(C_\uparrow= \exists x (A(x)\supset \forall y A(y))\)\\
   (2) $\quad $ \(C_\downarrow= \exists x (\exists y A(y)\supset  A(x))\).
\end{center}
 \(C_\downarrow\)
 is valid in all \(G_m\), 
\(G_\uparrow\). 
On the other hand,
\(C_\uparrow\) is valid in all \(G_m\) and 
\(G_\uparrow\) but not in \(G_\downarrow\). However, none of them is valid in \(G_{[0,1]}\) [\cite{Baaz-Leitsch-Zach:FirstOrderGoedel}, Corollary
2.9]. 
\end{proof}
\begin{remark}\label{r - 1.2}
   The formulas  \(C_\uparrow\) and \(C_\downarrow\) are important in the study of first-order infinite valued Gödel logics.  \(C_\uparrow\) expresses the fact that every infimum 
   is a minimum, while \(C_\downarrow\) states that every supremum 
   is a maximum.
\end{remark}


\begin{proposition}\label{p - 1.3}
\(G_{[0,1]}=\bigcap_V G_V\) where \(V\) ranges over all Gödel sets. \(G_{[0,1]}\) logic is contained in every Gödel logic. So it is an intersection of all Gödel logics.    
\end{proposition} 
\begin{proof}
If $\Gamma \entails_V A$ for every G\"{o}del set $V$\!, then it does so in
  particular for $V = [0, 1]$.  Conversely, if $\Gamma \nentails_V A$ for a
  G\"{o}del set~$V$\!, there is a $V$\!-interpretation $\I$ with $\I(\Gamma) >
  \I(A)$. Since $\I$ is also a $[0,1]$-interpretation, $\Gamma \nentails_{[0,1]}
  A$. 
\end{proof}

\end{section}

\section{Gluing Argument}\label{2}
In this section, we present a crucial tool to show that $G_{[0,1]}$ does not admit logically equivalent prenex normal form whenever $0$ is not an isolated point, which can be extended to all cases when 0 is isolated from the truth value set. 
So we literally cut the truth set at the point $\omega$, we keep the lower part and we glue the upper part to 1.
In this way, we simply change interpretations on atoms; The ones below  \(\omega\) remain and those above $\omega$ will become $1$. In principle, it does not matter what value \(\omega\) we fix, but if $\omega$ is not isolated from the truth values set the construction does not work for the universal quantifier.

\begin{lemma}[Gluing lemma for formulas with \(\exists\)]\label{l - 2.1}
Let \(\mathcal{I}\) be an interpretation into \(V\subseteq[0,1]\). Let us fix a value \(\omega\in[0,1]\) and define 
\begin{center}
$ \mathcal{I}_\omega(\mathcal{P} ) = \begin{cases}\mathcal{I}(\mathcal{P}) & \text{if\/ $\mathcal{I}(\mathcal{P} )\leq \omega$,} \\
               1     & \text{otherwise}\end{cases} $
\end{center}
for atomic formula \(\mathcal{P}\) in \(\mathcal{L}^\mathcal{I}\). Then \(\mathcal{I}_\omega\)
is an interpretation into \(V\) such that 
\begin{center} 
 $ \mathcal{I}_\omega(\mathcal{B} ) = \begin{cases}\mathcal{I}(\mathcal{B}) & \text{if\/ $\mathcal{I}(\mathcal{B} )\leq \omega$,} \\
               1     & \text{otherwise}\end{cases} $
\end{center}
  
\end{lemma} 
\begin{proof}
The proof is by induction principle after the complexity of the formulas that any subformula \(\mathcal{B}\) of \(\mathcal{A}\) with $\mathcal{I}(\mathcal{B} )\leq \omega$ has $\mathcal{I}'(\mathcal{B} )=\mathcal{I}(\mathcal{B} )$. For \(\mathcal{B}\) being an atomic formula it is obvious by definition. 
We distinguish cases according to the logical form of \(\mathcal{B}\):\\
$\mathcal{B} \equiv D \land E$.  If $\I(\mathcal{B}) \le \omega$, then, without loss of
    generality, assume
    $\I(\mathcal{B}) = \I(D)\le \I(E)$.  By induction hypothesis,
    $\I'(D) = \I(D)$ and $\I'(E) \ge \I(E)$, so $\I'(\mathcal{B}) = \I(\mathcal{B})$. If
    $\I(\mathcal{B})>\omega$, then $\I(D)>\omega$ and $\I(E)>\omega$, by induction
    hypothesis $\I'(D)=\I'(E)=1$, thus, $\I'(\mathcal{B})=1$.\\
    $\mathcal{B} \equiv D \lor E$.  If $\I(\mathcal{B}) \le v$, then, without loss of
    generality, assume
    $\I(\mathcal{B}) = \I(D) \ge \I(E)$.  By induction hypothesis,
    $\I'(D) = \I(D)$ and $\I'(E) = \I(E)$, so $\I'(\mathcal{B}) = \I(\mathcal{B})$.
    If $\I(\mathcal{B})>\omega$, then, again without loss of generality, $\I(\mathcal{B})=\I(D)>\omega$, by induction hypothesis $\I'(D)=1$, thus,
    $\I'(\mathcal{B})=1$.\\
    $\mathcal{B} \equiv D \impl E$.  Since $\omega < 1$, we must have
    $\I(D) > \I(E) = \I(\mathcal{B})$. By induction hypothesis,
    $\I'(D) \ge \I(D)$ and $\I'(E) = \I(E)$, so $\I'(\mathcal{B}) = \I(\mathcal{B})$.
    If $\I(\mathcal{B})>\omega$, then $\I(D)\ge\I(E)=\I(\mathcal{B})>\omega$, by induction
    hypothesis $\I'(D)=\I'(E)=\I'(\mathcal{B})=1$.\\
    $\mathcal{B} \equiv \qe xD(x)$. First assume that $\I(\mathcal{B})\le \omega$. Since $D(c)$
    evaluates to a value less or equal to $\omega$ in $\I$ and, by
    induction hypothesis, in $\I'$ also the supremum of these values
    is less or equal to $\omega$ in $\I'$, thus $\I'(\mathcal{B}) = \I(\mathcal{B})$.
    If $\I(\mathcal{B})>\omega$, then there is a $c$ such that $\I(D(c))>\omega$, by
    induction hypothesis $\I'(D(c))=1$, thus, $\I'(\mathcal{B})=1$.    
\end{proof}
The gluing lemma holds in general for all formulas iff the fixed point $\omega$ is isolated from the truth set from above. 
\begin{corollary}[Gluing lemma for for formulas with \(\forall\)]
 If \(\omega\notin Val(\mathcal{I}, \mathcal{B})\), thus \(\omega\) is isolated point from above, then the gluing lemma can be extended to the universal quantifier.
\end{corollary} 

\begin{proof}
Let    $\mathcal{B} \equiv \qa xD(x)$. This is the crucial part, since the above gluing argument has one proviso, it does not work for \(\forall\) if  \(\omega\) is not isolated from above. First, assume that
    $\I(\mathcal{B}) < \omega$. Then there is a witness~$c$ such that
    $\I(\mathcal{B}) \le \I(D(c)) < \omega$ and, by induction hypothesis, also
    $\I'(D(c))<\omega$ and therefore, $\I'(\mathcal{B}) = \I(\mathcal{B})$. For $\I(\mathcal{B}) > \omega$ it is
    obvious that $\I'(\mathcal{B}) = \I(\mathcal{B}) = 1$. Finally assume that
    $\I(\mathcal{B}) = \omega$. If this infimum would be proper, i.e.\ no
    minimum, then the value of all witnesses under $\I'$ would be $1$, but
    the value of $\mathcal{B}$ under $\I'$ would be $\omega$, which would contradict
    the definition of the semantic of the $\qa$ quantifier. Since all infima are minima, there is a witness $c$ such that
    $\I(D(c))=\omega$ and therefore, also $\I'(D(c))=\omega$ and thus
$\I'(\mathcal{B})=\I(\mathcal{B})$.
\end{proof}



As a consequence of the gluing lemma we have the following results:
\begin{proposition}\label{p - 2.1} 
In the following cases, 1-satisfiability in Gödel logics is classical satisfiability:\\
1) In the propositional case\\ 
2) In the first-order case, where the truth value set is arbitrary but 0 is isolated\\
3) In the case of the prenex fragment for any truth value set \\
4) In the case of the existential fragments for any truth value set
\end{proposition}
\begin{proof}
The first two cases follow by induction on the complexity of the formulas.
3) The change of values of the modified interpretations in the case of \(\omega\) being
non-isolated does not diminish the truth of prenex formulas. 
4) It follows from the fact that purely existential formulas are always classically satisfiable.
\end{proof}
We know that the gluing lemma works whenever there is no universal quantifier or the gluing point is isolated. Since in finite-valued Gödel logics the points are isolated, and so is in $G_\uparrow$, therefore we have the following result:
\begin{proposition}\label{p - 2.2}\cite{Baaz-Leitsch-Zach:FirstOrderGoedel}
\(G_\uparrow =\bigcap_{m\geq 2} G_m\) is an intersection of all
finite-valued Gödel logics.
\end{proposition}
\begin{proof}
The idea of the proof is the following: formula valid in $G_\uparrow$ is valid in all finite-valued Gödel logics. On the other hand, the formula not valid in $G_\uparrow$ is not valid in at least one finite-valued Gödel logic. Because the counterexample has a certain finite value, we glue everything from above this value to $1$, then we get a counterexample in finite-valued Gödel logic.\\
The right inclusion follows from proposition \ref{p - 1.2} 
It is sufficient to show that if \(G_\uparrow\nvDash A\) for $A$ being existential, then \(G_m\nvDash A\) for some $m$. Given any interpretation $\mathcal{I}$ and a fixed value $\omega < 1$, we define modified interpretation  
$\mathcal{I}_\omega$ and apply gluing lemma to the formula $A$. Now suppose that 
 \(\mathcal{I}\nvDash A\) then, clearly, if $A$ is an existential formula of the type $\exists \bar{x}B(\bar{x})$ then \(\mathcal{I}(A)= \omega< 1\) and there are only finitely many truth values below $\omega$ in $V$; let us say $\omega=1-1/k$. Since 
$\mathcal{I}(B)\leq \omega < 1$, by gluing lemma $\mathcal{I}_\omega(B)\leq \omega$ and so \(\mathcal{I}_\omega(A)\leq \omega\) but $\mathcal{I}_\omega$ is a $G_{k+1}$ interpretation and \(\mathcal{I}_\omega\nvDash A\).
\end{proof}

So here is our first result on the level of classification, specifically for $G_\downarrow$:
 \begin{proposition}\label{p - 2.3}
For all first-order Gödel logics, when $0$ is not an isolated point, formulas do not in generally allow the existence of an equivalent prenex normal form. 
 \end{proposition}
\begin{proof}
Consider the formula  $F:= \neg \forall x A(x)\wedge \forall x \neg \neg A(x)$ which states that 'for a suitable choice of $A(x)$, $0$ is a true infimum'. Suppose there exists an equivalent prenex normal form for some quantifier-free $P$ and some string of quantifiers $Qx'$
\begin{align*}
   F \Leftrightarrow  Qx' Px'.
\end{align*}
We can find a suitable interpretation of $A(x)$ (descending to $0$ but never reaching it) that makes the formula $F$ true. As a consequence, we have that the formula $F$ is $1$-satisfiable. Since the equivalence is valid, the prenex normal formula is also $1$-satisfiable. Moreover prenex formula is classically satisfiable so the gluing argument always works but the original formula $F$ can not be classically satisfiable, which leads to a contradiction.
\end{proof}


\begin{remark}\label{r - 2.1}
As a consequence, we have that, firstly, $1$-unsatisfiability is the same for all Gödel logics where $0$ is an isolated point. 
So the refutation of the $1$-satisfiability 
is recursive, because it is recursive in classic logic. The same is true for the existential fragments. 
\end{remark}



 \section{ \texorpdfstring{$>0$} - Valid Equivalent Prenex Normal Forms}\label{3}
The gluing argument can be used to solve another problem, namely if every double negated formula is equivalent to a double negated prenex normal form, in this case, classification is simple.\\
Recall that
\begin{proposition}\label{p - 3.1}
 1-validity of the double negated formula $\neg\neg A$ is equivalent to $>0$-validity of $A$.  
\end{proposition}
\begin{proof}
If we take the formula $A$ which has a value $>0$, then $\neg A$ has truth value 0, and therefore $\neg\neg A$ evaluates to 1. On the other hand, if we take $\neg\neg A$ which is evaluated to 1, then $\neg A$ has truth value 0, and therefore $A$ has value $>0$.
\end{proof}
 
 \begin{proposition}\label{p - 3.2}
If $0$ is not isolated in $V$, the first-order G\"odel logic \(G_V\)  does not allow in general the existence of the $>0$ valid equivalent prenex normal form.    
 \end{proposition}

\begin{proof}
Suppose that every formula admits its $>0$ valid equivalent prenex formula. Recall that $>0$ valid is expressed by double negation.
Let us define a formula and its prenex normal form such that the equivalence is $>0$ valid:
 \begin{center}
     \(\neg \neg (\neg \forall x A(x)\wedge \forall x \neg \neg A(x) \Leftrightarrow  Qx' Px')\).
 \end{center}
The proof is similar to the proof of proposition \ref{p - 2.2}, the only difference is that instead of $1$-satisfiable we consider $>0$ satisfiable. 
\end{proof}
\begin{proposition}\label{p - 3.3}
For all first-order Gödel logics when $0$ is an isolated point, every formula allows equivalent prenex normal forms such that the equivalence is $>0$-valid.
\end{proposition}
 \begin{proof}
Before we start the proof, recall that intuitionistic predicate logic $\IL$ is contained in all first-order Gödel logics, and all the Intuitionistically sound axioms and rules hold in \(G_V\). We will show that for every formula, its double negation is valid in Gödel logic when 0 is an isolated point, using Glivenko translation + Kuroda shift. One straightforward way is to consider Hilbert-style calculus with Modus-Ponens for the Gödel logic.\\
It is an easy observation that all axioms and rules of $\IL$ are intuitionistically sound with double negation.
For any formula \(A\), we have \(A\rightarrow \neg \neg A\) 
holds in $\IL$, it means that if we have assumption \(A\) true, then we can derive \(\neg \neg A\). Also, the axiom of the excluded middle with double negation \(\neg \neg(\neg A\vee A)\) holds in $\IL$. 
By induction, we show that every classically valid $A$ not including $\forall$ can be derived \(\neg \neg A\) in $\IL$. This is because Modus Ponens with double negation also holds
\begin{align*}
   ( \mathrm{\neg \neg MP}) & \quad \vcenter{\infer{\neg \neg B}{\neg \neg(A\rightarrow B) & \neg \neg A}}
\end{align*}
and 
\begin{align*}
 \hbox{($\neg \neg$Q$\exists$)} &\quad \vcenter{\infer{(\neg \neg\qe xA(x))\limp B^{(x)}}{\neg \neg(A(x))\limp B^{(x)}}}
\intertext{(where $B^{(x)}$ means that $x$ is not free in $B$).}   
\end{align*}
Now for the $\forall$
\begin{align*}
    \hbox{($\neg \neg$Q$\forall$)} &\quad \vcenter{\infer{B^{(x)}\limp(\neg \neg\qa xA(x))}{B^{(x)}\limp \neg \neg(A(x))}}
\end{align*}
this is obtained by Glivenko translation 
 \begin{align*}     \vcenter{\infer{\forall x\neg \neg A(x)}{\neg \neg (t)}}.\end{align*}
 
 adding  Kuroda shift (by simply applying double negation before the whole formula and after \(\forall\))
 \begin{align*}      \vcenter{\infer{\neg \neg\forall x A(x)}{\forall x \neg \neg A(x)}}.  
 \end{align*}
Note that \(\forall x\neg \neg A(x)\rightarrow \neg \neg\forall x  A(x)\) is $\axiso$ and is satisfied by assumption.
Thus, we have shown that for every classically valid formula, its double negation is valid in Gödel logic when 0 is an isolated point, this is obtained by Glivenko translation + Kuroda shift. We also know that classically, every formula is equivalent to a prenex normal form. Consequently, the double negation of the equivalence is also valid in Gödel logic, since it corresponds to $>0$ validity of the equivalence, which ends our proof.  
 \end{proof}

So by gluing argument, we have shown that no ($>0$ valid) equivalent prenex normal form exists, if $0$ is not an isolated point in the truth value set. 
It is natural to ask whether it is true in general that whenever 0 is isolated, the equivalent prenex normal form exists.
From now on, we investigate the existence of the equivalent prenex normal form for the first-order Gödel logics according to the number and type of accumulation points in their truth set.

\section{At Least One Accumulation Point From Above}\label{4}
In this section, we consider the first-order Gödel logics with at least one accumulation point from above in their truth value set. 
The idea is to reflect the truth values above the accumulation point $X$ and associate it with the logic \(G_{[X,1]}\).
We look at it as a subinterval, where we fix the value for this lower (accumulation) point. Then, of course, we can find an isomorphism to the full interval from this subinterval, which contains only this stretched upper part, under the condition that everything is interpreted above this accumulation point and nothing is interpreted below. 
\begin{theorem}\label{t - 4.1}
In first-order Gödel logics $G_V$ such that at least one accumulation point from above exists, there is no valid equivalent prenex normal form. 
\end{theorem}
\begin{proof}
The idea of the proof is that we introduce for the lower accumulation point from above a new propositional variable $X$ that behaves like 'falsum' in the usual sense. We reflect the truth values above $X$ and associate it with the logic \(G_{[X,1]}\). Then we may extend the interval $[X,1]$ to $[0,1]$ interval and apply the result from the previous section.
Let us consider the old formula 
\begin{center}
(1)  \(\neg \forall x A(x)\wedge  \forall x \neg \neg A(x)\).
\end{center}
and modify it in such a way that nothing has to be interpreted below $X$. 
Now we write the following literal interpretation of (1) by adding atomic formulas and adding \(X\) to every atomic formulas 
\begin{center}
 (2)  \(F\equiv (\forall x( A(x)\vee X)\supset X)\wedge  (\forall x ((A(x)\vee X )\supset X)\supset (A(x)\vee X))\).    
 \end{center}
Let \(Qx' Px'\) be the equivalent prenex normal form
 \begin{center}
 \(F\Leftrightarrow Qx' Px'\). 
 \end{center}

Then we know that this prenex normal form has to be evaluated above \(X\), since everything has value above \(X\). 
However, the matrix of the prenex formula might contain the $\bot$, therefore we replace them with a new 'falsum' \(X\) and apply the gluing argument (this does not change the evaluation, since the prenex formula is interpreted above $X$).
There is a suitable interpretation of $A(x)$ such that the formula $F$ is true. Therefore, it is 1-satisfiable, and so is the prenex formula. We can glue everything over $X$ to $1$ in the prenex formula. Then $X$ behaves as usual 'Falsum', so the prenex formula has a 2-valued model, where the values are the evaluation of $X$ and 1. However the original formula $F$ can not have such 2-valued model.
\end{proof}

\section{ No Accumulation Point From Above }\label{5}

Now we consider first-order G\"odel logics with truth set where all accumulation points are from below and there is no accumulation point from above.
We will distinguish the following different subcases: \begin{enumerate}
    \item No accumulation points exist at all (the finite case)
    \item Only one accumulation point from below at $1$ exists ( \(G_\uparrow\) )
    \item Only accumulation points from below exist but at least one is an inner accumulation point, which is less than $1$.
\end{enumerate} 
Prenexation is, of course, feasible in cases where all classical quantifier shift rules are
available. This holds for all finitely valued first-order Gödel logics and their intersection
\(G_\uparrow\) (the only infinite-valued Gödel logic where every formula is equivalent to a prenex normal form).
\begin{proposition}\label{p - 5.1}
 All finite first-order Goedel logics \(G_m\) admit equivalent prenex normal form.
\end{proposition}
\begin{proof}
 All finite-valued first-order G\"odel logics \(G_m\) admit the classical quantifier-shift rules, including the three quantifier-shift rules which are not valid in intuitionistic logic
  \begin{align*}
  (S1)  \quad  \forall x(A(x)\vee B)\supset(\forall xA(x) \vee B) \\
(S2)   \quad ( B\supset \exists xA(x))\supset(\exists x B\supset \vee A(x))\\
(S3) \quad  (\forall x A(x)\supset B)\supset(\exists x(A(x) \supset B)  
\end{align*}
(\(x\) is not free in \(B\))
  This is easily seen by considering the validity of the following formulas 
\begin{center}
    (1) $\quad $ \(C_\uparrow= \exists x (A(x)\supset \forall y A(y))\)\\
   (2) $\quad $ \(C_\downarrow= \exists x (\exists y A(y)\supset  A(x))\).
\end{center}
The formulas  \(C_\uparrow\) and \(C_\downarrow\) are important in the study of first-order infinite valued Gödel logics.  \(C_\uparrow\) expresses the fact that every infimum 
   is a minimum, while \(C_\downarrow\) expresses that every supremum 
   is a maximum, consequently they are valid \(G_m\).
(S1) is valid in all Gödel logic,
\((S2)\) is equivalent to \(C_\uparrow\) and \((S3)\) is equivalent to \(C_\downarrow\).
Consequently, all finite-valued first-order Gödel logics \(G_m\)  admit logically equivalent prenex normal forms. 
\end{proof}
\(G_\uparrow\) is thus the only infinite-valued Gödel logic where every formula
is equivalent to a prenex formula.
\begin{proposition}\label{p - 5.2}
\(G_\uparrow\) admits logically equivalent prenex normal forms.   
\end{proposition}
\begin{proof}
We have shown that $G_\uparrow$ is the intersection of all finite first-order Goedel logics.
Consequently, all classical quantifier shifting rules hold in  $G_\uparrow$.
\end{proof}

For the third subcase, we use a completely different approach.  In the previous discussion, we have defined a formula that describes the fact that the objects descend to $0$ but never reach. 
Now we do the opposite. We define a formula that describes a value is not an accumulation point in the interpretation 
and we consider, unlike the previous cases, models of various sizes. Note that to argue in that way we use the property that there is no accumulation point from above, because of this we are allowed usual model-theoretic skolemization for universal quantifiers.
\(\forall x A(x)\limp A(f)\) is trivial, for the other direction we need to introduce skolem axioms
\(A(f)\limp\forall x A(x)\) which is allowed since the infimum is always a minimum.

\begin{proposition}\label{p - 5.3}
For all first-order G\"odel logics that have only accumulation points from below such that at least one accumulation point is inside this interval, no equivalent prenex normal form exists.
\end{proposition}

\begin{proof}
 


Step 1:
Let us describe what it means "\(X\) not being a true accumulation point" for a suitable interpretation of atoms \(X\) and \(A(x)\)  assuming that $X<1$:
\begin{center}
$F\equiv\quad$\(\forall x(A(x)< X)\supset \exists x A(x)< X\) 
\end{center}
Step 2:
Assume there is a valid equivalent prenex normal form $Qx' Px'$. Then this prenex normal form, for all interpretations of \(X\) and \(A(x)\) should hold.\\
Step 3: Concentrate on interpretations under which all $A(x)<X<1$. Show that among these interpretations, there are some under which the equivalence is false.\\ 
Step 4: Since, by assumption, there is no accumulation point from above we add skolem axioms to\footnote{It is like in classic logic one can simply pick an element from above as there is no true infimum, 'every infimum is minimum'.},  skolemize $\qa$ on the side of the prenex form and reduce it to a purely existential formula.\\
Step 5:
Consider an $1$-element model. 
For any choice of \(X\) and \(A(x)\) the formula $F$ is always true.\\
Step 6: Extend the $1$-element model by adding an ascending sequence of points never reaching $X$ in such a way that the formula $F$ becomes false, the existential formula remains true. 
We add points, and at the same time, we evaluate the skolem axiom. We always take the minimum among the points that make the skolem axioms true. This is possible since there is no accumulation point from above. \\
But by assumption, this equivalence has to hold for all interpretations of \(X\) and \(A(x)\), so it should be $1$-satisfiable for the interpretation when all \(A(x)\) are below \(X\). But the formula $F$ might not be true, since for a suitable choice of ascending sequence of \(A(x)\) there might be an inner accumulation point. Thus, the equivalence does not hold. $F$ will not be true at accumulation point because accumulation point always take the value among the \(A(x)\)'s
and \(A(x)<1\), because all \(A(x)<X\), but the supremum of \(\exists x A(x)\) is not less then X, but X.
\end{proof}

\section{ Gödel logics with Delta operator \texorpdfstring{\(G_V^\Delta\)}{}}\label{6}
In this chapter, we investigate first-order Gödel logics with the absoluteness operator $\del$ after the existence of the equivalent prenex normal form. Recall that 
\begin{align*}    
 \I(\del A) = \begin{cases} 1 & \text{if $\I(A)=1$,}\\ 0 &\text{otherwise}\end{cases}
\end{align*}
This absoluteness
operator allows to represent the complement of $1$-satisfiability by
$\lnot\del A$, and to represent the complement of ${>}0$-satisfiability
by $\lnot A$ (note that $\lnot\lnot A$ means that~$A$ has value $>0$,
and therefore $\lnot\lnot\lnot A$ represents the complement of
${>}0$-satisfiability, and $\lnot\lnot\lnot A$ reduces to $\lnot A$
already intuitionistically.)
It is shown in \cite{Baaz-Preining:classificationoffirstorder} that -- contrary to classical logic -- there is
no faithful and uniform (across all truth value sets) recursive
translation which reduces the validity of first-order Gödel sentences
to the complement of $1$-satisfiability. 
\begin{remark}\label{r - 6.1}
 1-satisfiability of formulas with $\Delta$ does not imply classical satisfiability. For example the formula $\neg\neg A\land \neg\del A$ which says that $A$ is between 0 and 1, is 1-satisfiable with $\del$ but not classically 1-satisfiable.      
\end{remark}
Now we will answer the main question of this section: 'Can there be a prenex normal form with $\del$ when there is none without $\del$? '\\
We formulate our argument in the restricted semantics, where all atoms besides $\top$ are interpreted below 1. The point of restricted semantics is that we can evaluate $\del$ (because the value of $\del X$ for some atom $X$ which admits a value below 1 is 0). Therefore any formula with $\del$ coincides with the disjunction of chains of a formula without $\del$. 
It is clear that all counterexamples work for restricted semantics. We show that counterexamples work also if the prenex formulas contain $\del$.\\
Let us first recall some important notions we will use in this section:
\begin{definition}\label{d - 6.1}
 A \emph{chain} $C$ over the set of propositional variables $X=[X_1,...X_n]$ is an expression 
 \begin{align*}
(\bot \vartriangleright_1 X_{\Pi(1)})\land (X_{\Pi(1)}\vartriangleright_2 X_{\Pi(2)})\land\dots \land(X_{\Pi(n)}\vartriangleright_{n+1} \top)
        \end{align*}
abbreviated as 
\begin{align*}
\bot \vartriangleright_1 X_{\Pi(1)}\vartriangleright_2
        \top \vartriangleright_1 X_{\Pi(2)}\vartriangleright_3\dots \vartriangleright_n X_{\Pi(n)}\vartriangleright_{n+1} \top
        \end{align*}
    where $\vartriangleright_i\in \{<, \lequiv\}$ and $\Pi$ is permutation on $[1\dots n]$.
\end{definition}
Every chain describes an order type of the variables $X$.
We define the two types of chain normal forms which we will use later.
\begin{definition}\label{d - 6.2}($CNF_\del-1$).
The first \emph{chain normal form with $\del$} is the following expression
 \begin{align*}
     \lor C\land \psi_c(A)
 \end{align*}
where $\lor C$ are all chains and $\psi_c(A)$ is an evaluation\footnote{The evaluation changes subformula step by step according to the information contained in the chain.} over the chain $C$ of $A$ with variables among the variables of the chain. 
\end{definition}


\begin{definition}\label{d - 6.3}($CNF_\del-2$)
The second \emph{chain normal form with $\del$} in the restricted semantics is obtained from the first chain normal form with $\del$ after the following steps:\\
1) If the evaluation of $A$ is false we delete the chain,\\
2) if the evaluation of $A$ is true we leave the chain as it is,\\
3) if the evaluation of $A$ is an atom and is not 1 we delete the whole chain.
\end{definition}
\begin{lemma}\label{l - 6.1}
The second chain normal form of a formula with $\del$ in the restricted semantics coincides with a disjunction of chains without $\del$.  
\end{lemma}
\begin{proof}
Let us take a formula $F$ without $\del$ and develop the second chain normal form.
According to Definition 6.3 the disjuncts might be among the following three cases:\\
1) $C \land \top$, \\
2) $C \land \bot$ or \\
3) $C \land X$ for some atom $X$ and chain $C$. \\
Since we are in the restricted semantics, all atoms are interpreted below 1. Now we put $\del$ in front of the disjuncts and we distribute.
3) becomes false because atoms can not be interpreted at 1 so the evaluation of $\del X$ is 0 and we remove it. 2) will be removed by definition and therefore remains only 1) which is equivalent to $\del C$. 
In the restricted semantics, $\del C$ coincides with $C$, thus the second chain normal form will coincide with the disjunction of chains without $\del$.    
\end{proof}

As a consequence of this lemma the prenex formulas with 
$\del$ coincides with the prenex formulas without $\del$ in the restricted semantics.
\begin{theorem}\label{t - 6.1}
There is no equivalent prenex normal form with $\del$ for the formulas in the first-order G\"odel logics with $\del$ when there is none for the formulas in the first-order G\"odel logics without $\del$.
\end{theorem}
\begin{proof}
Suppose that the formula without prenex normal form without $\del$ has prenex normal form with $\del$. Then the equivalence holds also in the restricted semantics.
The matrix of the prenex formula corresponds to the disjunction of chains.
From the above lemma follows that any formula 
with $\del$ coincides with the disjunction of chains of the formula without $\del$. In restricted semantics adding $\del$ to the chains does not change anything. this contradicts to our assumption.
\end{proof}
\begin{itemize}
\item Thus, all the negative results from the first-order G\"odel logics without $\del$ can be transferred to first-order G\"odel logics with $\del$.
\item There are only two positive cases to investigate.
 \end{itemize}
 It is easy to see that the gluing argument is not valid in all first-order G\"odel logics with \(\del\).
\begin{proposition}\label{p - 6.1}
Finite-valued G\"odel logics with $\del$ still admit equivalent prenex normal form. \end{proposition}
\begin{proof}
 By the following two equivalences, 
 \begin{align*}
\del \forall x A(x)\lequiv \forall x \del A(x)\\
\del \exists x A(x)\lequiv \del A(x)   
   \end{align*}
one can shift quantifiers, so in the finite case with $\del$, all quantifier shifting rules are valid, and therefore equivalent prenex normal form exists.   
\end{proof}
\begin{proposition}\label{p - 6.2}
$G_\uparrow^\del$ is not an intersection of finite valued G\"odel logics.    
\end{proposition}
\begin{proof}
 Since in finite-valued logics also with $\del$ all quantifier-shifts rules are valid, but not in $G_\uparrow^\del$ (e.g., $\del \exists x A(x)\lequiv \del$ A(x) is not valid), therefore \(G_\uparrow^\del\) is not an intersection of finite G\"odel logics anymore. 
\end{proof}
\begin{proposition}\label{p - 6.3}
$G_\uparrow^\Delta$ admits no equivalent prenex normal form. 
\end{proposition}
\begin{proof}
The proof is similar to Proposition 5.3. 
In $G_\uparrow^\del$, we formulate an expression 
that takes value either true or false
\begin{center}
  $F\equiv\forall x \neg \del A(x)\rightarrow \neg \del\exists x A(x)$.
\end{center}
The formula says that '1 is not an accumulation point'; More precisely, the formula is false at 1 when 1 is the accumulation point.
Assuming this formula has an equivalent prenex normal form, it can also be evaluated as true or false. 
we can use the skolem axioms 
\begin{center}
    $A(f)\rightarrow \forall x A(x)$
\end{center}
to reduce universal quantifiers to skolem functions, we are allowed to do so since there are no accumulation points from above and it can not happen that the infimum is empty.
With the skolem axioms, we reduce the prenex formula to the purely existential formula.
We know the formula is true and thus the existential formula is also true in a $1$-element model. It is true in the bigger model too whereas $F$ is not true for a suitable choices of interpretation. 
\end{proof}

\begin{lemma}\label{p - 6.4}
In any G\"odel logic, given an interval $(0,c]$ such that no atoms are interpreted in this interval, they can be interpreted only at 0 or above $ c$, then it follows that every formula is either interpreted at 0 or $\geq c$. 
\end{lemma}
\begin{proof}
The proof is simply by induction. For atoms, it is stated; For propositional connectives, it is clear due to G\"odel conditional. For $\forall$ there might exist a proper infimum which is $c$. For  $\exists$, since all elements are interpreted bigger then $c$, then the supremum is also bigger then $c$.     
\end{proof}

\begin{theorem}\label{t - 6.2}
In first-order infinite-valued G\"odel logics with $\del$ when 0 is an isolated point, in general there exists no $>0$-valid equivalent prenex normal form.
\end{theorem}
\begin{proof}
Suppose there is an accumulation point from above.\\
Step 1: Consider the expression 'not being an accumulation point from above' 
 \begin{align*}
F\equiv (\forall x\del (X< A(x))\limp \del(X<\forall xA(x))))
 \end{align*}
 This formula has the property that it evaluates either to true or false.\\
 Step 2: Since the surrounding of 0 is isolated, we fix some lower point $c$ of the surrounding of 0. Consider the interpretation where the surrounding of 0 is empty and all $A(x)$ are interpreted above $c$ or is 0. Thus, the infimum might be empty if $A(x)$ is not interpreted 0. Recall that we are in restricted semantics, where nothing is interpreted between 0 and $c$.\\
Step 3: Suppose $F$ admits $>0$-valid equivalent prenex form
 \begin{align*}
 \neg\neg(F\lequiv Qx' Px'). 
 \end{align*}
As everything besides 0 is interpreted above or equal $c$, we can find an interpretation of the skolem functions such that the skolem axioms are interpreted above $c$. Consequently, there is a $>0$ valid equivalence between the formula $F$ and existential prenex formula.\\
Step 4: Repeat the same argument as in Proposition \ref{p - 5.3}. just in this case the equivalence is ($>c>0$)-valid. Consider a 1-element model, the equivalence is $>0$ in the mentioned interpretation.
Now we extend the 1-element model by adding a descending chain of points that converges to $X$ but never reaches $X$, in such a way that the existential formula remains $>0$-valid, and we make sure that the skolem axioms remain $>0$ valid by simply adding the lower element for the infimum which is always $\geq c$ if it is not 0.\\
 Step 5: However, at the accumulation point within the interval $(0,c]$, the formula $F$ becomes 0 for a suitable interpretation in the restricted semantics, but the prenex formula remains $>0$-valid. So the equivalence is 0 and not $>0$. \\
In case there are no accumulation points from above we consider the following formula:
 \begin{align*}
F\equiv  \forall x\del(A(x)<X)\limp \del(\exists x A(x)<X).
 \end{align*}
 We repeat the same argument as above.
\end{proof}

\begin{corollary}\label{c - 6.2.1}
There is no $>0$-valid equivalent prenex normal form for formulas in general in $G_\uparrow^\del$.
\end{corollary}

\section{Validity Equivalence}\label{7}
In this section, we consider a weaker form of prenexation in G\"odel logics namely that the validity of a formula is effectively\footnote{Can be constructed one to one recursive procedure that assigns equivalent valid prenex formula to a given original formula.} equivalent to the validity for some prenex normal form. (This question is senseless if the equivalence is admitted to be non-effective: take $\bot$ and $\top$, respectively). \\
In this case e.g. $\gdl {[0,1]}$ admits prenex forms as derivability in $\gdl {[0,1]}$ can be expressed
by Kleene’s T\footnote{It is a universal Turing machine for classical logic that expresses the fact that the formula can be derived iff it is in the recursively enumerable set.} within $\gdl {[0,1]}$ putting double negation in front of all atoms and shifting
the quantifiers in the classical way.\\
This normal form does not always exist as is demonstrated by the fact that $G_V$
with $V$ uncountable but $0$ not in the perfect subset is not recursively enumerable but
its prenex fragment is.\\
The problem of the effective existence of validity (1-satisfiability)
equivalent prenex forms can be sharpened to the effective existence of validity (1-
satisfiability) equivalent purely existential\footnote{Note that in the case of validity, universal quantifiers are replaced. In the case of
1-satisfiability existential quantifiers are replaced.} (universal) forms.

\begin{proposition} \label{p - 7.1} The following holds in first-order G\"odel logic:\\
 1. Prenex formulas admit validity equivalent standard skolemizations,\\
2. Prenex formulas admit 1-satisfiability equivalent standard skolemizations.   
\end{proposition}
\begin{proof}
1) It is sufficient to prove for an arbitrary formula $F$ and a new function $f$ the following:
\begin{align*}
    \models \exists\Bar{x}\forall y F(\Bar{x}, y) \Leftrightarrow \models \exists\Bar{x} F(\Bar{x}, f(\Bar{x})).
\end{align*}
The proof is by induction. The direction from left to right is obvious.\\
For the other direction, suppose that $\nvDash \exists\Bar{x}\forall y F(\Bar{x}, y) $ then for some interpretation $\mathcal{I}\omega$ 
 \begin{align*}
     sup\{\mathcal{I}\omega(\forall y F(\Bar{c}, y))=\omega_{\Bar{c}}\}\leq \omega<1.
 \end{align*}
Using the axiom of choice we can assign a value for every $f(\Bar{c})$ such that $\mathcal{I}_\omega(F(\Bar{c}, f(\Bar{c})))$ is in between $\omega_{\Bar{c}}$ and $\omega_{\Bar{c}}+ (1-\omega)/2$. As a consequence 
\begin{align*}
     sup\{\mathcal{I}\omega(F(\Bar{c}, f(\Bar{c}))) < (1-\omega)/2\}\}\leq \omega+(1-\omega)/2<1. 
 \end{align*}
Thus the right is not valid.\\
2) It is obvious since 1-satisfiability coincide with classical satisfiability.
\end{proof}
\begin{remark}\label{r - 7.1}
 As a consequence a prenex formula is valid if the corresponding existential formula is valid.
\end{remark}
Validity in first-order G\"odel logic is characterized by the following theorem.
\begin{theorem}\label{t - 7.1}\cite{Baaz-Preining:classificationoffirstorder}
A first-order G\"odel logic \(G_V\) is recursively enumerable iff one of the following conditions is satisfied:\\
1. \(V\) is finite,\\
2. \(V\) is uncountable\footnote{A G\"odel set is uncountable iff it contains a non-trivial dense linear subordering.}, and $0$ is an isolated point,\\
3. \(V\) is uncountable, and every neighbourhood of $0$ is uncountable.
\end{theorem}
In all the remaining cases, i.e., logics with countable truth value set and those with uncountable truth value set where
there is a countable neighbourhood of $0$ and $0$ is not isolated; the respective
logics are not r.e.
\begin{remark}\label{r - 7.2}
Note that finitely valued logics of different cardinalities differ and finitely valued
logics of the same cardinality coincide. 2) and 3) in the above theorem characterize single logic each.
3) is axiomatized by the axioms and rules of intuitionistic predicate calculus
extended by (Lin) and (QS).
2) is axiomatized by the addition of $Iso_0$ axiom to 3).
 In case  $V$ has a finite
cardinality \(n\), then 1) is axiomatized by adding 
the respective axiom Fin(n) to 3).
\end{remark}

The following propositions characterize the G\"odel logics with uncountable G\"odel set:
\begin{proposition}\label{p - 7.2}
 The only G\"odel logics with an uncountable set that admit validity equivalent prenex normal forms are in case when $0$ is isolated or  $0$ is in perfect set. 
\end{proposition}
\begin{proof}
 The reason is that the prenex fragment in the uncountable case is always recursively enumerable.
 It was shown by \cite{Baaz-Preining:classificationoffirstorder}
 that when $0$ is isolated or  $0$ is in perfect set it is recursively enumerable. On the other hand, when every surrounding of  $0$ is countable ( $0$ is not in the perfect set),
the set is not recursively enumerable but the prenex fragment is. Therefore there is no equivalent valid prenex formula since there is no effective recursive construction. Note that if the logic is r.e. Kleene's T can be used to express the formulas and Kleene's T can be represented in Goedel logics by adding n double negation to every atom. This enables classical quantifier shifting rules.
\end{proof} 
\begin{remark}\label{r - 7.3}
Note that all non-recursively enumerable G\"odel logics allow the expression of validity
in all finite domains.
\end{remark}

\begin{proposition}\label{p - 7.3}
Finite valued Gödel logics and $G_\uparrow$ admits validity equivalent prenex normal forms.
\end{proposition}
\begin{proof}
In all cases where there exists an equivalent prenex form by quantifier shifts, there is also an effective validity equivalent prenex normal forms, therefore finite valued G\"odel logics and $G_\uparrow$ also have effective validity equivalent prenex forms.    
\end{proof}
 

\section{Full Classification Tables}\label{8}

The results of this paper can be summarized in the following tables:
\begin{table}[H]
\begin{tabular}{|l|l|l|l|l|}
\hline
\multicolumn{5}{|c|} {Logically Equivalent Prenex Normal Forns} \\ \hline
G\"odel set $V$ & without $\del$ & with $\del$ & validity $>0$ & Validity $>0$ with $\del$ \\ \hline
finite & \checkmark & \checkmark & \checkmark & \checkmark \\ \hline
$G_\uparrow$ & \checkmark &\texttimes   & \checkmark &\texttimes   \\ \hline
Countable without $G_\uparrow$ & \texttimes  & \texttimes
 & \texttimes  & \texttimes
 \\ \hline
 0 isolated & \texttimes  & \texttimes
 & \checkmark  & \texttimes
 \\ \hline
 0 not isolated &\texttimes
  &\texttimes   &\texttimes
  &\texttimes   \\ \hline
\end{tabular}
\end{table}

\begin{table}[H]
\centering
\begin{tabular}{|llll|}
\hline
\multicolumn{4}{|c|}{Validity Equivalent Prenex Normal Forns} \\ \hline
\multicolumn{2}{|l|}{Goedel set $V$}& \multicolumn{1}{l|}{with $\del$} & without $\del$ \\ \hline
\multicolumn{2}{|l|}{Finite} & \multicolumn{1}{l|}{\checkmark } & \checkmark  \\ \hline
\multicolumn{1}{|l|}{\multirow{3}{*}{Uncountable}} & \multicolumn{1}{l|}{0 isolated} & \multicolumn{1}{l|}{\checkmark} & \checkmark  \\ \cline{2-4} 
\multicolumn{1}{|l|}{} & \multicolumn{1}{l|}{0 is in perfect set} & \multicolumn{1}{l|}{\checkmark} &\checkmark  \\ \cline{2-4} 
\multicolumn{1}{|l|}{} & \multicolumn{1}{l|}{0 not in perfect set} & \multicolumn{1}{l|}{\texttimes}  &\texttimes   \\ \hline
\multicolumn{2}{|l|}{Countable}   & \multicolumn{2}{|l|}{Open}   \\ \hline
\end{tabular}
\end{table}

\section{Further Research}\label{9}
For the countable case, one can ask whether there is an effectively constructible validity equivalence between formulas and existential formulas.
The existence of such prenex forms depends on reflective properties of the non-recursively
enumerable G\"odel logics. H\'ajek \cite{Hajek:non-arithmetical} has provided a non-arithmetic lower bound and
Aguilera, Byd\v zovsk\'y, Fern\'andez-Duque \cite{Juan:nonhyperarithmetical} have provided a corresponding upper bound
for $G_\downarrow$ ($G_\downarrow$ is the countable G\"odel logic with sole accumulation point at 0). 
An idea
would be the use of partial truth definitions. On the other hand the countability of the
logic might provide isolated points which provide a partition of the truth-value set.\\
We conjecture the following:\\
\begin{conjecture}\label{conject} A first-order G\"odel logic admits effective validity equivalent prenex normal form iff it is arithmetical.
\end{conjecture}
1) We can investigate whether purely existential prenex formulas form an arithmetical set. In this case we will obtain a full characterization. \\
2) We can study other G\"odel sets after the property of the logic being arithmetical.

\begin{remark}\label{r - 9.1}A logic being non-arithmetical does not mean that a single formulas are also non-arithmetical.\end{remark}




\bibliographystyle{plain}

\begin{thebibliography}{10}
\bibitem{Juan-Baaz:Unsound}Juan P Aguilera and Matthias Baaz.
\newblock Unsound inferences make proofs shorter.
arXiv preprint arXiv:1608.07703, 2016.

\bibitem{Juan:nonhyperarithmetical} Juan Pablo Aguilera, Jan Byd\v zovsk\'y, and David Fern\'andez-Duque. 
\newblock  A nonhyperarithmetical
G\"odel logic. 
\newblock In International Symposium on Logical Foundations
of Computer Science, pages 1–8. Springer, 2022.


\bibitem{Baaz:Delta}
Matthias Baaz.
\newblock Infinite-valued {G}{\"{o}}del logic with 0-1-projections and
  relativisations.
\newblock In Petr H{\'{a}}jek, editor, {\em {G}{\"{o}}del'96: {L}ogical
  Foundations of Mathematics, Computer Science, and Physics}, volume~6 of {\em
  Lecture Notes in Logic}, pages 23--33. Springer-Verlag, Brno, 1996.

\bibitem{Baaz-Ciabattoni:TaitCutElimGoedel}
Matthias Baaz and Agata Ciabattoni.
\newblock A {S}ch{\"{u}}tte-{T}ait style cut-elimination proof for first-order
  {G}\"{o}del logic.
\newblock In Uwe Egly and Christian~G. Ferm{\"{u}}ller, editors, {\em Automated
  Reasoning with Analytic Tableaux and Related Methods, International
  Conference, TABLEAUX 2002}, volume 2381 of {\em LNCS}, pages 24--38, Berlin,
  2002. Springer.

\bibitem{Baaz-Ciabattoni-Fermuller:GodelSurvey}
Matthias Baaz, Agata Ciabattoni, and Christian~G. Ferm{\"{u}}ller.
\newblock Hypersequent calculi for {G}{\"{o}}del logics: A survey.
\newblock {\em Journal of Logic and Computation}, 13(6):835--861, 2003.

\bibitem{Baaz-Ciabattoni-Fermuller:MonadicFragments}
Matthias Baaz, Agata Ciabattoni, and Christian~G. Ferm{\"{u}}ller.
\newblock Monadic fragments of {G}{\"{o}}del logics: Decidability and
  undecidability results.
\newblock In Nachum Dershowitz and Andrei Voronkov, editors, {\em Logic for
  Programming, Artificial Intelligence, and Reasoning}, volume 4790/2007 of
  {\em Lecture Notes in Computer Science}, pages 77--91, 2007.

\bibitem{Baaz-Ciabattoni-Preining:SATborderline}
Matthias Baaz, Agata Ciabattoni, and Norbert Preining.
\newblock {SAT} in monadic {G}{\"{o}}del logics: {A} borderline between
  decidability and undecidability.
\newblock In Hiroakira Ono, Makoto Kanazawa, and Ruy {de Queiroz}, editors,
  {\em Logic, Language, Information and Computation, 16th Workshop, WoLLIC
  2009}, number 5514 in LNAI, pages 113--123, 2009.

\bibitem{Baaz-Ciabattoni-Preining:SatGoedel}
Matthias Baaz, Agata Ciabattoni, and Norbert Preining.
\newblock First-order satisfiability in {G}{\"{o}}del logics: An {NP}-complete
  fragment.
\newblock {\em Theoretical Computer Science}, 412:6612--6623, 2011.

\bibitem{Baaz-Ciabattoni-Zach:QuantifiedPropositionalGoedel}
Matthias Baaz, Agata Ciabattoni, and Richard Zach.
\newblock Quantified propositional {G}{\"{o}}del logic.
\newblock In Andrei Voronkov and Michel Parigot, editors, {\em Logic for
  Programming and Automated Reasoning. 7th International Conference, LPAR
  2000}, volume 1995 of {\em LNAI}, pages 240--256, Berlin, 2000. Springer.

\bibitem{Baaz-Fermueller:ResolutionPrenexGoedel}
Matthias Baaz and Christian~G. Ferm{\"{u}}ller.
\newblock A resolution mechanism for prenex {G}\"{o}del logic.
\newblock In Anuj Dawar and Helmut Veith, editors, {\em Computer Science Logic,
  24th International Workshop, CSL 2010, 19th Annual Conference of the EACSL},
  volume 6247 of {\em LNCS}, pages 67--79, Berlin, 2010. Springer.

\bibitem{Baaz-Leitsch-Zach:TimeGaps}
Matthias Baaz, Alexander Leitsch, and Richard Zach.
\newblock Completeness of a first-order temporal logic with time-gaps.
\newblock {\em Theoretical Computer Science}, 160(1--2):241--270, 1996.

\bibitem{Baaz-Leitsch-Zach:FirstOrderGoedel}
Matthias Baaz, Alexander Leitsch, and Richard Zach.
\newblock Incompleteness of a first-order {G}{\"{o}}del logic and some temporal
  logics of programs.
\newblock In {\em Computer Science Logic}, volume 1092/1996 of {\em Lecture
  Notes in Computer Science}, pages 1--15, 1996.

\bibitem{Baaz-Preining:QElimOmegaKripke}
Matthias Baaz and Norbert Preining.
\newblock Quantifier elimination for quantified propositional logics on
  {K}ripke frames of type omega.
\newblock {\em Journal of Logic and Computation}, 18:649--668, 2008.

\bibitem{Baaz-Preining:Gödel–Dummett}
Matthias Baaz, Norbert Preining, \newblock Gödel–Dummett logics, \newblock in: Petr Cintula, Petr Hájek, Carles Noguera (Eds.) {\em Handbook of Mathematical Fuzzy Logic} vol.2, College Publications, 2011, pp.585–626, chapterVII.

\bibitem{Baaz-Preining:classificationoffirstorder} Matthias Baaz and Norbert Preining. 
\newblock On the classification of first order Goedel logics. 
\newblock {\em  Ann. Pure Appl. Log} 170:36–57, 2019.


\bibitem{Baaz-Preining-Zach:GoedelFragments}
Matthias Baaz, Norbert Preining, and Richard Zach.
\newblock Characterization of the axiomatizable prenex fragments of first-order
  {G}\"{o}del logics.
\newblock In {\em 33rd IEEE International Symposium on Multiple-Valued Logic
  (ISMVL 2003)}, pages 175--180, Los Alamitos, 2003. IEEE Computer Society.

\bibitem{Baaz-Preining-Zach:CompletenessGoedelDelta}
Matthias Baaz, Norbert Preining, and Richard Zach.
\newblock Completeness of a hypersequent calculus for some first-order
  {G}\"{o}de logics with delta.
\newblock In {\em 36th International Symposium on Multiple-valued Logic (ISMVL
  2006)}. IEEE Computer Society, 2006.

\bibitem{Baaz-Preining-Zach:FirstOrderGoedel}
Matthias Baaz, Norbert Preining, and Richard Zach.
\newblock First-order {G}{\"{o}}del logics.
\newblock {\em Annals of Pure and Applied Logic}, 147(1--2):23--47, 2007.

\bibitem{Baaz-Veith:QuantifiedPropositionalGodel}
Matthias Baaz and Helmut Veith.
\newblock An axiomatization of quantified propositional {G}{\"{o}}del logic
  using the {T}akeuti-{T}itani rule.
\newblock In {\em Logic Colloquium 1999}, volume~13 of {\em Lecture Notes in
  Logic}, pages 91--104, 2000.

\bibitem{Baaz-Veith:Interpolation}
Matthias  and Helmut Veith. 
\newblock Interpolation in Fuzzy Logic
\newblock {\em Archive for Mathematical Logic},
accepted for publication.

\bibitem{Baaz-Veith:QuantifierElimination}
Matthias  and Helmut Veith.
\newblock Quantifier Elimination in Fuzzy Logic
\newblock CSL 1998.



\bibitem{Baaz-Zach:Compact}
Matthias Baaz and Richard Zach.
\newblock Compact propositional {G}{\"{o}}del logics.
\newblock In {\em Proceedings of 28th International Symposium on
  Multiple-Valued Logic}, pages 108--113, Los Alamitos, CA, 1998. IEEE Computer
  Society Press.

\bibitem{Baaz-Zach:HypersequentIntuitionisticFL}
Matthias Baaz and Richard Zach.
\newblock Hypersequents and the proof theory of intuitionistic fuzzy logic.
\newblock In Peter~G. Clote and Helmut Schwichtenberg, editors, {\em
  Proceedings of 14th {CSL} Workshop}, volume 1862 of {\em Lecture Notes in
  Computer Science}, pages 187--201, Berlin, 2000. Springer-Verlag.

 \bibitem{Buechi:Onadecisionmethod}
 J.R Buechi.
\newblock On a decision method in restricted second order arithmetic.
\newblock In  {\em Logic,
Methodology, and Philosophy of Science }
Proc. of the 1960 Congress, Stanford University
Press, pp. 1-11.


\bibitem{Beckmann-Goldstern-Preining:Fraisse}
Arnold Beckmann, Martin Goldstern, and Norbert Preining.
\newblock Continuous {F}ra{\"{\i}}ss{\'{e}} conjecture.
\newblock {\em Order}, 25(4):281--298, 2008.

\bibitem{Beckmann-Preining:LinearKripkeGoedel}
Arnold Beckmann and Norbert Preining.
\newblock Linear {K}ripke frames and {G}{\"o}del logics.
\newblock {\em Journal of Symbolic Logic}, 72(1):26--44, 2007.


\bibitem{Beckmann-Preining:Decidinglogics}
Arnold Beckmann, Norbert Preining, \newblock Deciding logics of linear Kripke frames with scattered end pieces,
\newblock {\em Soft Comput}. 21(1) (2017) 191–197.

\bibitem{Beckmann-Preining:Separatingintermediate}
Arnold Beckmann, Norbert Preining, \newblock Separating intermediate predicate logics of well-founded and dually well-founded structures by monadic sentences,
\newblock  {\em J. Logic Comput}. 25(3) (2015) 527–547.



\bibitem{Ciabattoni:GlobalIL}
Agata Ciabattoni.
\newblock A proof-theoretical investigation of global intuitionistic (fuzzy)
  logic.
\newblock {\em Archive for Mathematical Logic}, 44(4):435--457, 2005.

\bibitem{Davey:Subvarieties}
Brian~A. Davey.
\newblock On the lattice of subvarieties.
\newblock {\em Houston Journal of Mathematics}, 5(2):183--192, 1979.

\bibitem{Dummett:GodelLogic}
Michael Dummett.
\newblock A propositional calculus with denumerable matrix.
\newblock {\em Journal of Symbolic Logic}, 24(2): 97--106, 1959.

\bibitem{Fitting:IntuitionisticForcing}
Melvin~Chris Fitting.
\newblock {\em Intuitionistic Logic, Model Theory and Forcing}.
\newblock Studies in Logic and the Foundations of Mathematics. North-Holland,
  Amsterdam/London, 1969.

\bibitem{Fitting:ProofMethods}
 Melvin~Chris Fitting.
\newblock Proof Methods for Modal and Intuitionistic Logics. 
\newblock Kluwer, 1983.


\bibitem{Fraisse:Ordres}
Roland Fra{\"{\i}}ss{\'e}.
\newblock Sur la comparaison des types d'ordres.
\newblock {\em Comptes Rendus Hebdomadaire des S{\'{e}}ances de
  l'Acad{\'{e}}mie des Sciences, Paris}, 226:1330--1331, 1948.

\bibitem{Godel:ZumAussagen}
Kurt G{\"{o}}del.
\newblock Zum intuitionistischen {A}ussagenkalk{\"{u}}l.
\newblock {\em Anzeiger Akademie der Wissenschaften Wien}, 69: 65--66, 1932.

 \bibitem{Gottwald:Mehrwertige}
 S. Gottwald.
\newblock Mehrwertige Logik.
\newblock Berlin 1989.


\bibitem{Hajek:1998}
Petr H{\'{a}}jek.
\newblock {\em Metamathematics of Fuzzy Logic}, volume~4 of {\em Trends in
  Logic}.
\newblock Kluwer, Dordrecht, 1998.

\bibitem{Hajek:non-arithmetical}
Petr H\'ajek. A non-arithmetical G\"odel logic. \newblock {\em Logic Journal of the IGPL}, 13(4):435–441, 2005.



\bibitem{Horn:LinearHeyting}
Alfred Horn.
\newblock Logic with truth values in a linearly ordered {H}eyting algebras.
\newblock {\em Journal of Symbolic Logic}, 34(3):395--408, 1969.

\bibitem{Kechris:DescriptiveSetTheory}
Alexander~S. Kechris.
\newblock {\em Classical Descriptive Set Theory}, volume 159 of {\em Graduate
  Texts in Mathematics}.
\newblock Springer, 1995.

 \bibitem{Kreiser-Gottwald-Stelzner:Nichtklassische}
L. Kreiser, S. Gottwald, W. Stelzner  
\newblock Nichtklassische Logik.
\newblock {\em Akademie-Verlag},
Berlin 1988.


 \bibitem{Maksimova:Craig}
 L. Maksimova.
\newblock Craig’s interpolation theorem and amalgamable varieties.
\newblock {\em  Doklady
Akademii Nauk SSSR}, 237/6, 1281-1284, 1977.


\bibitem{Mansfield:DefinableSets}
Richard Mansfield.
\newblock Perfect subsets of definable sets of real numbers.
\newblock {\em Pacific Journal of Mathematics}, 35:451--457, 1970.

\bibitem{Moschovakis:DescriptiveSetTheory}
Yiannis~N. Moschovakis.
\newblock {\em Descriptive Set Theory}, volume 100 of {\em Studies in Logic and
  the Foundations of Mathematics}.
\newblock North-Holland, Amsterdam, 1980.

\bibitem{Ono:KripkeModels}
Hiroakira Ono.
\newblock Kripke models and intermediate logics.
\newblock {\em Publications of the Research Institute for Mathematical Sciences
  Kyoto University}, 6:461--476, 1971.

\bibitem{Ono:IntermediatePredicate}
Hiroakira Ono.
\newblock A study of intermediate predicate logics.
\newblock {\em Publications of the Research Institute for Mathematical Sciences
  Kyoto University}, 8:619--649, 1972/73.

\bibitem{Preining:GoedelAxiomatizability}
Norbert Preining.
\newblock {\em Complete Recursive Axiomatizability of {G}{\"{o}}del Logics}.
\newblock PhD thesis, Vienna University of Technology, Austria, 2003.

\bibitem{Preining:Cantor–Bendixon}
Norbert Preining, 
\newblock Gödel logics and Cantor–Bendixon analysis, 
\newblock in: M. Baaz, A. Voronkov (Eds.), {\em Logic for Programming and Automated Reasoning}, LPAR 2002. Proceedings, in: Lecture Notes in Artificial Intelligence, vol.2514, Springer, 2002, pp.327–336.


\bibitem{Raney:CDCLattices}
George~N. Raney.
\newblock Completely distributive complete lattices.
\newblock {\em Proceedings of the American Mathematical Society}, 3:677--680,
  1952.

\bibitem{Rosenstein:LinearOrderings}
Joseph~G. Rosenstein.
\newblock {\em Linear Orderings}.
\newblock Academic Press, New York, 1982.

\bibitem{Takano:StrongCompletenessIFL}
Mitio Takano.
\newblock Another proof of the strong completeness of the intuitionistic fuzzy
  logic.
\newblock {\em Tsukuba Journal of Mathematics}, 11(1):101--105, 1987.

\bibitem{Takano:RQKripkeModels}
Mitio Takano.
\newblock Ordered sets {R} and {Q} as bases of {K}ripke models.
\newblock {\em Studia Logica}, 46:137--148, 1987.

 \bibitem{Takeuti:ProofTheory}
 Gaisi Takeuti
\newblock Proof Theory, 
\newblock 2nd ed., {\em North-Holland} , 1987.

\bibitem{Takeuti-Titani:Intuitionistic}
Gaisi Takeuti and Satoko Titani.
\newblock Intuitionistic fuzzy logic and intuitionistic fuzzy set theory.
\newblock {\em Journal of Symbolic Logic}, 49(3):851--866, 1984.

\bibitem{Troelstra:ConstructiveMathematics}
Anne~S. Troelstra.
\newblock Aspects of constructive mathematics.
\newblock In Jon Barwise, editor, {\em Handbook of Mathematical Logic}, pages
  973--1052. North-Holland, 1977.

\bibitem{Winkler:OrderTopology}
Reinhard Winkler.
\newblock How much must an order theorist forget to become a topologist?
\newblock In {\em Contributions of General Algebra 12, Proc. of the Vienna
  Conference}, pages 420--433, Klagenfurt, Austria, 1999. Verlag Johannes Heyn.





\end{thebibliography}



\end{document}